\PassOptionsToPackage{unicode}{hyperref}
\PassOptionsToPackage{hyphens}{url}
\PassOptionsToPackage{dvipsnames,svgnames,x11names}{xcolor}
\documentclass[
]{article}
\usepackage{amsmath,amssymb}
\usepackage{lmodern}
\usepackage{iftex}
\ifPDFTeX
  \usepackage[T1]{fontenc}
  \usepackage[utf8]{inputenc}
  \usepackage{textcomp} % provide euro and other symbols
\else % if luatex or xetex
  \usepackage{unicode-math}
  \defaultfontfeatures{Scale=MatchLowercase}
  \defaultfontfeatures[\rmfamily]{Ligatures=TeX,Scale=1}
\fi
\IfFileExists{upquote.sty}{\usepackage{upquote}}{}
\IfFileExists{microtype.sty}{% use microtype if available
  \usepackage[]{microtype}
  \UseMicrotypeSet[protrusion]{basicmath} % disable protrusion for tt fonts
}{}
\makeatletter
\@ifundefined{KOMAClassName}{% if non-KOMA class
  \IfFileExists{parskip.sty}{%
    \usepackage{parskip}
  }{% else
    \setlength{\parindent}{0pt}
    \setlength{\parskip}{6pt plus 2pt minus 1pt}}
}{% if KOMA class
  \KOMAoptions{parskip=half}}
\makeatother
\usepackage{xcolor}
\usepackage{graphicx}
\makeatletter
\def\maxwidth{\ifdim\Gin@nat@width>\linewidth\linewidth\else\Gin@nat@width\fi}
\def\maxheight{\ifdim\Gin@nat@height>\textheight\textheight\else\Gin@nat@height\fi}
\makeatother
\setkeys{Gin}{width=\maxwidth,height=\maxheight,keepaspectratio}
\makeatletter
\def\fps@figure{htbp}
\makeatother
\NewDocumentCommand\citeproctext{}{}
\NewDocumentCommand\citeproc{mm}{%
  \begingroup\def\citeproctext{#2}\cite{#1}\endgroup}
\makeatletter
 \let\@cite@ofmt\@firstofone
 \def\@biblabel#1{}
 \def\@cite#1#2{{#1\if@tempswa , #2\fi}}
\makeatother
\newlength{\cslhangindent}
\newlength{\csllabelwidth}
\newenvironment{CSLReferences}[2] % #1 hanging-indent, #2 entry-spacing
 {\begin{list}{}{%
  \setlength{\itemindent}{0pt}
  \setlength{\leftmargin}{0pt}
  \setlength{\parsep}{0pt}
  \ifodd #1
   \setlength{\leftmargin}{\cslhangindent}
   \setlength{\itemindent}{-1\cslhangindent}
  \fi
  \setlength{\itemsep}{#2\baselineskip}}}
 {\end{list}}
\usepackage{calc}

\ifLuaTeX
\usepackage[bidi=basic]{babel}
\else
\usepackage[bidi=default]{babel}
\fi
\babelprovide[main,import]{american}
\def\languageshorthands#1{}
\ifLuaTeX
  \usepackage{selnolig}  % disable illegal ligatures
\fi
\IfFileExists{bookmark.sty}{\usepackage{bookmark}}{\usepackage{hyperref}}
\IfFileExists{xurl.sty}{\usepackage{xurl}}{} % add URL line breaks if available
\hypersetup{
  pdftitle={SlicerNNInteractive: A 3D Slicer extension for
nnInteractive},
  pdfauthor={Coen de Vente, Kiran Vaidhya Venkadesh, Bram van Ginneken,
Clara I. Sánchez},
  pdflang={en-US},
  colorlinks=true,
  linkcolor={Maroon},
  filecolor={Maroon},
  citecolor={Blue},
  urlcolor={Blue},
  pdfcreator={LaTeX via pandoc}}

\title{SlicerNNInteractive: A 3D Slicer extension for nnInteractive}

\definecolor{c53baa1}{RGB}{83,186,161}
\definecolor{c202826}{RGB}{32,40,38}

\usepackage[affil-it]{authblk}
\usepackage{orcidlink}
\author[1,2%
  ]{Coen de Vente%
    \,\orcidlink{0000-0001-5908-8367}\,%
    }
\author[3%
  ]{Kiran Vaidhya Venkadesh%
    \,\orcidlink{0000-0002-4846-9049}\,%
    }
\author[3%
  ]{Bram van Ginneken%
    \,\orcidlink{0000-0003-2028-8972}\,%
    }
\author[1,2%
  ]{Clara I. Sánchez%
    \,\orcidlink{0000-0001-9787-8319}\,%
    }

\affil[1]{Quantitative Healthcare Analysis (qurAI) Group, Informatics
Institute, University of Amsterdam, Amsterdam, The Netherlands%
  }
\affil[2]{Amsterdam UMC location University of Amsterdam, Biomedical
Engineering and Physics, Amsterdam, The Netherlands%
  }
\affil[3]{Diagnostic Image Analysis Group (DIAG), Department of
Radiology and Nuclear Medicine, Radboud UMC, Nijmegen, The Netherlands%
  }
\date{7 April 2025}

\begin{document}
\maketitle

\section{Summary}\label{summary}

\texttt{SlicerNNInteractive} integrates \texttt{nnInteractive}
(\citeproc{ref-isensee2025nninteractive}{Isensee et al., 2025}), a
state-of-the-art promptable deep learning-based framework for 3D image
segmentation, into the widely used \texttt{3D\ Slicer} platform. Our
extension implements a client-server architecture that decouples
computationally intensive model inference from the client-side
interface. Therefore, \texttt{SlicerNNInteractive} eliminates heavy
hardware constraints on the client-side and enables better operating
system compatibility than existing plugins for \texttt{nnInteractive}.
Running both the client and server-side on a single machine is also
possible, offering flexibility across different deployment scenarios.
The extension provides an intuitive user interface with all interaction
types available in the original framework (point, bounding box,
scribble, and lasso prompts), while including a comprehensive set of
keyboard shortcuts for efficient workflow.

\section{Statement of Need}\label{statement-of-need}

Segmentation is a cornerstone of medical image analysis. Recently,
\texttt{nnInteractive} (\citeproc{ref-isensee2025nninteractive}{Isensee
et al., 2025}), a deep learning-based framework allowing for fast,
promptable segmentation of 3D medical images was released and was shown
to substantially outperform existing approaches, such as SAM2
(\citeproc{ref-kirillov2023segment}{Kirillov et al., 2023}), SegVol
(\citeproc{ref-du2024segvol}{Du et al., 2024}), and SAM-Med-3D
(\citeproc{ref-wang2023sam}{Wang et al., 2023}). Alongside the
\texttt{nnInteractive} model, plugins in the medical image viewers MITK
(\citeproc{ref-MITK_Team_MITK_2024}{MITK~Team, 2024}) and Napari
(\citeproc{ref-Sofroniew2025-ty}{Sofroniew et al., 2025}) were
published. However, the original authors did not make an extension
available for \texttt{3D\ Slicer}, a widely used viewer and processing
environment in medical imaging research. Furthermore, these existing
plugins require substantial computational resources on the machine of
the image viewer itself (an NVIDIA GPU with at least 10 GB of VRAM is
recommended), as these plugins do not facilitate the deployment of the
backend on a separate server. Moreover, \texttt{nnInteractive} only runs
on Windows and Linux, so the image viewer cannot be run on MacOS
machines.

\texttt{SlicerNNInteractive} decouples the computationally intensive
\texttt{nnInteractive} inference by allowing users to configure a remote
server (e.g., a node of a GPU cluster), while running the client on a
machine with lower computational capabilities. This approach not only
broadens platform compatibility, but also addresses the resource
constraints of existing plugins, making \texttt{nnInteractive} more
widely available and potentially accelerating research related to
promptable segmentation.

\section{\texorpdfstring{Overview of
\texttt{SlicerNNInteractive}}{Overview of SlicerNNInteractive}}\label{overview-of-slicernninteractive}

\subsection{nnInteractive}\label{nninteractive}

While foundation models such as SAM (\citeproc{ref-ravi2024sam}{Ravi et
al., 2024}) and SAM2 (\citeproc{ref-kirillov2023segment}{Kirillov et
al., 2023}) have shown promising interactive segmentation performance in
2D natural images, their lack of volumetric awareness and the domain
shift from natural to medical data resulted in limited utility in 3D
medical imaging contexts. \texttt{nnInteractive} addresses these issues
through an nnUNet-based architecture
(\citeproc{ref-isensee2021nnu}{Isensee et al., 2021}) with residual
encoders (\citeproc{ref-isensee2024nnu}{Isensee et al., 2024}) that
supports diverse interation types: point, bounding box, scribble, and
lasso prompts. Trained on over 120 diverse volumetric datasets across
multiple modalities (CT, MRI, PET, 3D microscopy), the framework
demonstrated high accuracy and versatility. Our implementation extends
this capability to \texttt{3D\ Slicer}.

\subsection{Availability and
Installation}\label{availability-and-installation}

\texttt{SlicerNNInteractive} is available through multiple channels. The
server-side is available through Docker Hub
(\texttt{docker\ pull\ coendevente/nninteractive-slicer-server:latest}),
Pip (\texttt{pip\ install\ nninteractive-slicer-server}), and GitHub
(\url{https://github.com/coendevente/nninteractive-slicer}). The
client-side is currently only available through our
\href{https://github.com/coendevente/nninteractive-slicer}{GitHub
repository}. In future versions of this extension, we plan to include it
in the official 3D Slicer Extensions Manager.

\subsection{Client-server Setup}\label{client-server-setup}

\begin{figure}
\centering
\includegraphics[width=0.8\textwidth,height=\textheight]{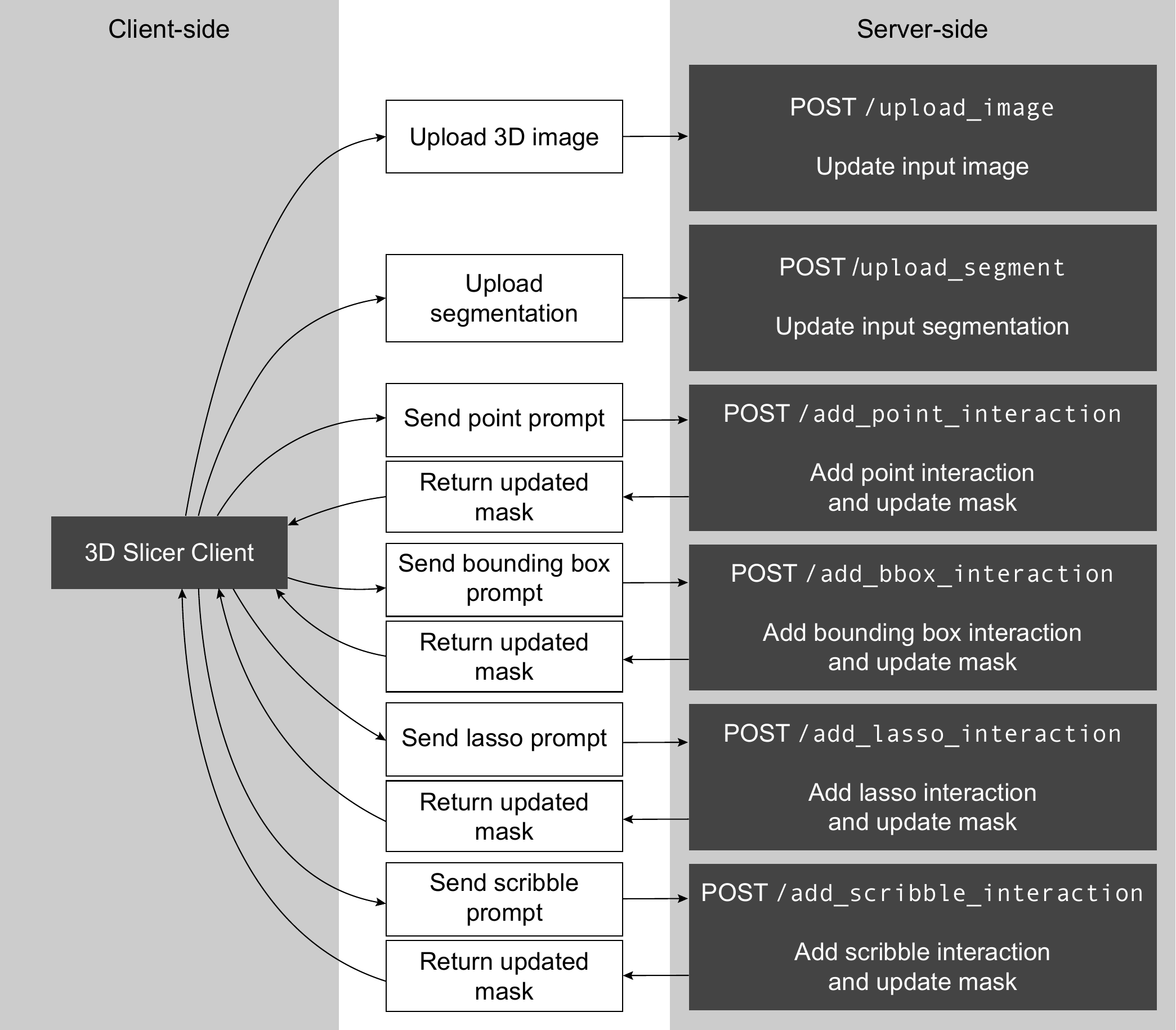}
\caption{API overview.\label{fig:api_overview}}
\end{figure}

\texttt{SlicerNNInteractive} uses a client-server setup, which decouples
the computationally intensive model inference from the
\texttt{3D\ Slicer} client. The server-side and client-side communicate
through FastAPI endpoints. The client maintains synchronization between
the image and input mask in \texttt{3D\ Slicer}. An overview of the API
is shown in \autoref{fig:api_overview}.

\subsection{User Interface}\label{user-interface}

\begin{figure}
\centering
\includegraphics{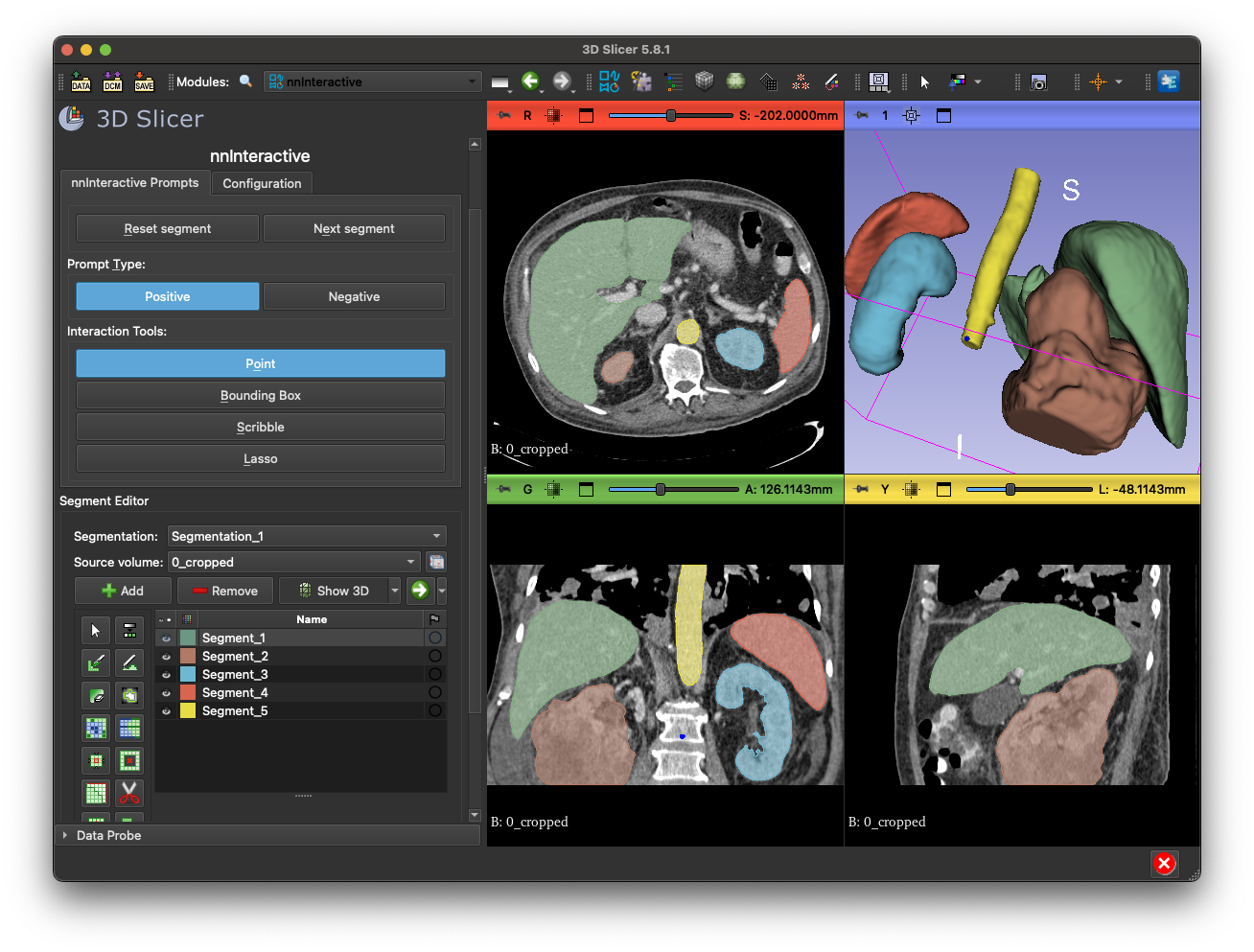}
\caption{Screenshot of the \texttt{SlicerNNInteractive}
extension.\label{fig:screenshot}}
\end{figure}

The user interface of \texttt{SlicerNNInteractive} largely follows the
\texttt{nnInteractive} Napari and MITK plugins. A screenshot of the user
interface, including segmentation results, is shown in
\autoref{fig:screenshot}. A video showcasing the functionalities of the
extension is available
\href{https://www.youtube.com/watch?v=mW_fUT1-IWM}{here}.

The sidebar of the user interface consists of a menu with the tabs
\emph{nnInteractive Prompts} and \emph{Configuration}, and the
\emph{Segment Editor}. The \emph{Configuration} tab allows the user to
change the Server URL. This URL is saved in \texttt{3D\ Slicer}'s
settings, which will be remembered in future sessions. The
\emph{nnInteractive Prompts} menu consists of the following sections:

\begin{itemize}
\item
  \textbf{Segment buttons:} The \emph{Reset segment} button removes all
  prompts from the current segment and deletes the current segmentation
  on the server and client-side. The \emph{Next segment} button creates
  a new empty segment in the \emph{Segment Editor}.
\item
  \textbf{Prompt Type:} These \emph{Positive} and \emph{Negative}
  buttons manage whether the provided prompt will be interpreted as a
  positive or negative prompt, respectively.
\item
  \textbf{Interaction Tools:} The four buttons in this section activate
  or deactivate the interaction tools. When a prompt type is activated,
  the user can place the prompt in the image. When a prompt has been
  placed, the client synchronizes the image and the segment to the
  server if needed, and sends the prompt to the server. The server
  subsequently processes the prompt and sends the updated segmentation
  back. When a prompt has been placed and processed, a new prompt of the
  same type can be placed immediately.
\end{itemize}

Each button in the \emph{nnInteractive Prompts} menu has an associated
keyboard shortcut, which is indicted using the underlined letters within
the button text.

If a segment is selected in the \emph{Segment Editor}, prompts will
always be applied to that segment. Every time a user has switched
segments, the associated segmentation is uploaded to server and used as
input mask to the \texttt{nnInteractive} model. When no segment is
selected, a new segment is created automatically.

\section*{References}\label{references}
\addcontentsline{toc}{section}{References}

\phantomsection\label{refs}


\begin{CSLReferences}{1}{0}
\bibitem[\citeproctext]{ref-du2024segvol}
Du, Y., Bai, F., Huang, T., \& Zhao, B. (2024). {SegVol}: Universal and
interactive volumetric medical image segmentation. \emph{Advances in
Neural Information Processing Systems}, \emph{37}, 110746--110783.

\bibitem[\citeproctext]{ref-isensee2021nnu}
Isensee, F., Jaeger, P. F., Kohl, S. A., Petersen, J., \& Maier-Hein, K.
H. (2021). {nnU-Net}: A self-configuring method for deep learning-based
biomedical image segmentation. \emph{Nature Methods}, \emph{18}(2),
203--211. \url{https://doi.org/10.1038/s41592-020-01008-z}

\bibitem[\citeproctext]{ref-isensee2025nninteractive}
Isensee, F., Rokuss, M., Krämer, L., Dinkelacker, S., Ravindran, A.,
Stritzke, F., Hamm, B., Wald, T., Langenberg, M., Ulrich, C., Deissler,
J., Floca, R., \& Maier-Hein, K. (2025). nnInteractive: Redefining 3D
promptable segmentation. \emph{arXiv Preprint arXiv:2503.08373}.
\url{https://doi.org/10.48550/arXiv.2503.08373}

\bibitem[\citeproctext]{ref-isensee2024nnu}
Isensee, F., Wald, T., Ulrich, C., Baumgartner, M., Roy, S., Maier-Hein,
K., \& Jaeger, P. F. (2024). {nnU-Net} revisited: A call for rigorous
validation in {3D} medical image segmentation. \emph{International
Conference on Medical Image Computing and Computer-Assisted
Intervention}, 488--498. \url{https://doi.org/10.48550/arXiv.2404.09556}

\bibitem[\citeproctext]{ref-kirillov2023segment}
Kirillov, A., Mintun, E., Ravi, N., Mao, H., Rolland, C., Gustafson, L.,
Xiao, T., Whitehead, S., Berg, A. C., Lo, W.-Y., Dollar, P., \&
Girshick, R. (2023). Segment anything. \emph{Proceedings of the IEEE/CVF
International Conference on Computer Vision}, 4015--4026.

\bibitem[\citeproctext]{ref-MITK_Team_MITK_2024}
MITK~Team. (2024). \emph{{MITK}} (Version v2024.12).
\url{https://github.com/MITK/MITK}

\bibitem[\citeproctext]{ref-ravi2024sam}
Ravi, N., Gabeur, V., Hu, Y.-T., Hu, R., Ryali, C., Ma, T., Khedr, H.,
Rädle, R., Rolland, C., Gustafson, L., Mintun, E., Pan, J., Alwala, K.
V., Carion, N., Wu, C.-Y., Girshick, R., Dollár, P., \& Feichtenhofer,
C. (2024). {SAM} 2: Segment anything in images and videos. \emph{arXiv
Preprint arXiv:2408.00714}.
\url{https://doi.org/10.48550/arXiv.2408.00714}

\bibitem[\citeproctext]{ref-Sofroniew2025-ty}
Sofroniew, N., Lambert, T., Bokota, G., Nunez-Iglesias, J., Sobolewski,
P., Sweet, A., Gaifas, L., Evans, K., Burt, A., Doncila Pop, D.,
Yamauchi, K., Weber Mendonça, M., Buckley, G., Vierdag, W.-M., Royer,
L., Can Solak, A., Harrington, K. I. S., Ahlers, J., Althviz Moré, D.,
\ldots{} Zhao, R. (2025). \emph{Napari: A multi-dimensional image viewer
for python}. Zenodo. \url{https://doi.org/10.5281/zenodo.8115575}

\bibitem[\citeproctext]{ref-wang2023sam}
Wang, H., Guo, S., Ye, J., Deng, Z., Cheng, J., Li, T., Chen, J., Su,
Y., Huang, Z., Shen, Y., \& others. (2023). {SAM-Med3D}: Towards
general-purpose segmentation models for volumetric medical images.
\emph{arXiv Preprint arXiv:2310.15161}.
\url{https://doi.org/arXiv.2310.15161}

\end{CSLReferences}
\end{document}